\documentstyle[11pt,epsf]{article}
\def\beq{\begin{eqnarray}}
\def\eeq{\end{eqnarray}}
\def\bea{\begin{eqnarray*}}
\def\eea{\end{eqnarray*}}

\def\centeron#1#2{{\setbox0=\hbox{#1}\setbox1=\hbox{#2}\ifdim
\wd1>\wd0\kern.5\wd1\kern-.5\wd0\fi
\copy0\kern-.5\wd0\kern-.5\wd1\copy1\ifdim\wd0>\wd1
\kern.5\wd0\kern-.5\wd1\fi}}
\def\ltap{\;\centeron{\raise.35ex\hbox{$<$}}{\lower.65ex\hbox{$\sim$}}\;}
\def\gtap{\;\centeron{\raise.35ex\hbox{$>$}}{\lower.65ex\hbox{$\sim$}}\;}

\def\singleandthirdspaced{\baselineskip=\normalbaselineskip\multiply
    \baselineskip by 130\divide\baselineskip by 100}

\newcommand{\newc}{\newcommand}
\newc{\qbar}{{\overline q}}
\newc{\Kahler}{K\"ahler }
\newc{\deltaGS}{\delta_{\rm GS}}
\begin{document}
\begin{titlepage}
\begin{flushright}
{\large hep-th/yymmnnn \\ SCIPP 12/16\\
}
\end{flushright}

\vskip 1.2cm

\begin{center}

{\LARGE\bf Discrete Symmetries/Discrete Theories}

\vskip 1.4cm

{\large Milton Bose and Michael Dine}
\\
{\it Santa Cruz Institute for Particle Physics and
\\ Department of Physics, University of California,
    Santa Cruz CA 95064  } \\
\vskip 4pt

\vskip 4pt

\vskip 1.5cm

\begin{abstract}
Dynamical, metastable supersymmetry breaking appears to be a generic phenomena in supersymmetric field theories.  It's simplest implementation is
within the so-called ``retrofitted O'Raifeartaigh Models".  While seemingly flexible, model building in these theories is significantly constrained.
In gauge-mediated versions, if the approximate $R$ symmetry of the theory is spontaneously broken, the messenger
scale
is fixed; if explicitly broken by retrofitted couplings, a very small dimensionless number is required; if supergravity corrections are
responsible for the symmetry breaking,
at least two moderately small couplings are required, and there is a large range
of possible messenger scales.  In gravity mediated versions, achieving $m_{3/2} \approx
M_Z$ is a problem of discrete tuning.  With plausible assumptions, one can't achieve this to better than a factor of $100$, perhaps accounting
for a {\it little hierarchy} and the surprisingly large value of the Higgs mass. 
\end{abstract}

\end{center}

\vskip 1.0 cm

\end{titlepage}
\setcounter{footnote}{0} \setcounter{page}{2}
\setcounter{section}{0} \setcounter{subsection}{0}
\setcounter{subsubsection}{0}

\singleandthirdspaced

\section{Introduction:  The Genericity of Metastable DSB}

As Nelson and Seiberg pointed out\cite{Nelson:1993nf}, generic, stable spontaneous supersymmetry breaking requires a continuous $R$ symmetry.  If we insist that 
there should be no exact continuous $R$ symmetries in nature, then we expect that, at some level, any continuous $R$ symmetry
should be explicitly broken, leading, generically, to restoration of supersymmetry somewhere in the space of fields.  Discrete symmetries,
on the other hand, are plausible in generally covariant theories, and indeed frequently arise in string constructions.\footnote{Whether they are ``typical",
and might emerge in a landscape context, is another question\cite{dinesun,Dine:2008jx,dinefestucciamorisse}.}  A simple possibility is that the
discrete symmetry is a subgroup of the required continuous $R$ symmetry.  This can readily be implemented to generate metastable O'Raifearataigh models.
For example, in a theory with fields $X,Y,A$ transforming, under a $Z_N$ symmetry, as:
\beq
X \rightarrow \alpha^2 X~~~~Y \rightarrow \alpha^2 Y~~~~ A \rightarrow A
\eeq
with $\alpha = e^{2 \pi i \over N}$, and also a $Z_2$ under which $A$ and $Y$ are odd, the superpotential has the structure
\beq
W = X (A^2 -  f) + m AY + \left ({Y A^3 \over M} + {X^{N+1}\over M^{N-2}} + \dots \right )
\label{ormodel}
\eeq
Ignoring the non-renormalizable couplings, the theory possesses a supersymmetry-breaking ground state at the origin of $X,Y$.  Including these couplings,
there is a supersymmetric ground state at large $X,Y$.  The supersymmetry-breaking state is metastable.  It exhibits
an approximate, continuous $R$ symmetry. This would seem a generic phenomenon.

One would like to understand the breaking of supersymmetry dynamically.  Models with stable dynamical supersymmetry breaking (DSB) were discovered
some time ago\cite{ads}; they seem quite special, and pose challenges for model building.  Models of
metastable DSB (MDSB) were considered by Intriligator, Shih and Seiberg\cite{iss} exhibited strongly coupled models which exhibit metastable dynamical supersymmetry
breaking.  The ISS class of models are a rich and interesting set of theories, but they pose challenges for building models.  An even broader
class of theories is obtained by studying the O'Raifeartaigh models, and rendering the scales ($f$ and $m$) in eqn. \ref{ormodel}, for example)
dynamical\cite{Dine:2006gm,Dine:2006xt,dinekehayias}.
 In these ``retrofitted"
models, the discrete $R$ symmetry is spontaneously broken by gaugino condensation or its generalizations\cite{dinekehayias}.  This symmetry breaking
can also readily generate a $\mu$ term.  If one retrofits an O'Raiferataigh model in which all fields have $R$ charge $0$ or $2$, one has a problem (also
typical of ISS models) that the approximate $R$ symmetry is not spontaneously broken.  A simple approach, adapted in \cite{dinekehayias}, is to retrofit one of the models of Shih\cite{Shih:2007av},
in which not all field have such charges.  But given the seeming freedom of
the retrofitted approach, it is interesting to ask whether one can break the continuous $R$ symmetry explicitly.   In particular, if there is a distinct, messenger sector, it would seem possible that retrofitting a breaking of the 
approximate $R$ symmetry might not spoil supersymmetry breaking.  This might allow construction of classes of models of
{\it General Gauge Mediation}\cite{Meade:2008wd}.  Alternatively, supergravity corrections might dominate, as has been discussed
by Kitano\cite{kitano}.  We'll see in this case
one can obtain the structure of minimal gauge mediation (MGM).

There is another interesting feature of the retrofitted models, stressed first in \cite{dinekehayias}.  If one assumes that higher dimension operators are controlled
by the Planck scale, $M_p$, then the expectation value of the superpotential, $\langle W \rangle$ is readily of the correct order of magnitude
to cancel the cosmological constant.  This is remarkable; it means that one neither has to introduce a peculiar, $R$-breaking constant in the superpotential,
nor introduce additional dynamics (e.g. additional gaugino condensates) to account for the observed dark energy (of course, one must still tune an order
one constant to incredible accuracy).  This is in contrast to the viewpoint, for example, of KKLT\cite{kklt}, that the constant in the superpotential is
to be thought of as a random number, selected as part of the anthropic determination of the cosmological constant.

If we insist on this relation, there are striking restrictions on the allowed theories.  We will see, in particular, that the underlying scale of
supersymmetry breaking (as measured by $m_{3/2}$), sometimes takes on discrete values.  In such theories, the usual questions
of fine tuning become a question of selection of discrete, rather than continuous, parameters.

In this note after reviewing generalized gaugino condensation in section \ref{review}, we briefly revisit the problem of retrofitting gravity mediation, focussing especially
on the discrete choices required (particularly in the sector responsible for discrete $R$ symmetry
breaking) in section \ref{gravitymediation}.  Here the observation concerning the cosmological constant relates the scale of the new interactions to $m_{3/2}$; with
some plausible assumptions about unification, this scale is determined, once one makes
a (discrete) choice of the underlying gauge group.  $m_{3/2}$ is then exponentially
dependent on the leading beta function of the underlying theory, and one can ask how closely one can (discretely)
tune the gravitino mass to $M_Z$.  We will see that with some plausible assumptions about coupling unification (more precisely,
a plausible {\it model} for coupling unification), one typically misses by factors of order $100$, perhaps providing an explanation
of a little hierarchy.

We then consider the problem of retrofitting models of gauge mediation in sections \ref{spontaneous}-\ref{supergravity}.
We will take the observation above about the cosmological constant as a guiding principle.  We will see that this is a significant constraint.
All of the models possess an approximate, continuous $R$ symmetry.  We will consider the possibilities that this symmetry is spontaneously
broken, or explicitly broken. Given the current experimental constraints, we will accept a significant degree of tuning, and take this scale to be large, of order $10^6$ GeV.  Tuned models of gauge mediation have been considered in \cite{fengetal}.

Apart from the fact that one can readily build realistic models, there are several striking features which emerge from these studies.
\begin{enumerate}
\item  In models with spontaneous breaking of the $R$ symmetry, the scalings are fixed by discrete choices.  Quite generally,
\beq
\sqrt{F} \approx 10^{9}~{\rm GeV}
\eeq
corresponding to a messenger scale of order $10^{12}$ GeV (an interesting number, for example, from the perspective of axion physics) and
$m_{3/2} = 1$ GeV.
\item  In models in which one retrofits an explicit breaking of the $R$ symmetry, small couplings are required in order that the graviton mass be small, and that the gauge-mediated
contributions dominate.
\item  In models in which the breaking of the $R$ symmetry arises from supergravity corrections (i.e. the low dimension terms in the theory
respect the $R$ symmetry), one can obtain acceptable models without exceptionally small dimensionless parameters.  The messenger
scale can range over a broad range of scales; in the simplest cases, the superparticle spectrum is that of mgm.
\item  As has been noted previously\cite{dinekehayias}, a suitable $\mu$ term can readily be obtained, though this typically requires the introduction of a small, dimensionless
number.
\item  As has been discussed elsewhere, if the $\mu$ term arises as a result of
retrofitting, $B_\mu$ is small, so $\tan \beta$ is large\cite{Dine:2006xt}.
\item  With the assumption of a large scale, $\Lambda_{gm}$, CP constraints are weakened.  In some of the models we will describe,
however, CP conservation is automatic.
\end{enumerate}
In section \ref{conclusions}, we present our conclusions.

\section{Brief Review of Discrete R Symmetries and Generalized Gaugino Condensation}
\label{review}

Crucial to most discussions of supersymmetry dynamics is gaugino condensation.  Gaugino condensation can be defined, in a general way,
as dynamical breaking of a discrete $R$ symmetry, accompanied by dimensional transmutation.  As such, it occurs in a wider variety of theories than
just pure (supersymmetric) gauge theories.  For example, an $SU(N)$ gauge theory with $N_f$ flavors, and a singlet, $S$, with superpotential
\beq
W = y_f S \bar Q_f Q_f + {\gamma \over 3}S^3
\eeq
has a $Z_{3N-N_f}$ $R$ symmetry.  This is broken by $\langle \lambda \lambda\rangle\sim 32 \pi^2 \Lambda^3$, and by
$\langle S \rangle$.  In the limit $\vert \gamma\vert \ll \vert y_f \vert$, $S$ is large, and one can integrate out
the quark fields, obtaining an effective superpotential:
\beq
W = N\left(\prod y_f\right)^{1/N}S^{N_f/N}\Lambda^{3-N_f/N} + {\gamma \over 3} S^3.
\eeq
This has supersymmetric stationary points with
\beq
S \sim\Lambda\left[\left(\prod y_f\right)^{1 \over N}{N_f\over\gamma}\right]^{N \over 3 N - N_f}
\eeq
(this model also has a disconnected, runaway branch; this can be avoided, if desired, by adding additional scalars).  
The low energy superpotential has a constant term,
\beq
W_0 = \langle -{1\over 4 g^2 }W_\alpha^2\rangle\sim N\Lambda^3
\eeq

With these ingredients we can readily ``retrofit" any O'Raifeartaigh model.  For example, we can take
\beq
W = X (A^2 - \mu^2) + m A Y
\eeq
and replace it by
\beq
W = X (A^2 - c {W_\alpha^2 \over M_p}) + \kappa S A Y.
\eeq
This model has a metastable minimum near the origin, as seen from the standard
Coleman-Weinberg calculation.  It has a runaway to a supersymmetric vacuum at $\infty$, separated by a barrier from the (metastable)
minimum at the origin.  
Under the discrete $R$, $X$ is neutral, while $A$ transforms like the gauginos, $S$ has charge $2/3$, and $Y$ charge $1/3$.
Various higher dimension terms are allowed, which lead to (faraway) supersymmetric vacua.

Clearly any dimensional coupling can be generated in this way, and the possibilities for model building are vast.
This type of construction will be the basic ingredient of all of the models of this paper.
One striking feature of this model is that, for $c$ an order $1$ constant, the cosmological constant can be very small;
upon coupling to supergravity, the terms $
\vert {\partial W \over \partial X} \vert^2
$
and $-{3 \over M_p^2} \vert W \vert^2$ are automatically of the same order of magnitude.
We view this remarkable coincidence as a potential clue, and will largely insist that it hold in the models we describe in this paper.
This will greatly restrict possibilities for model building.

\section{Retrofitted Gravity Mediation:  Discrete Choices}
\label{gravitymediation}

In gravity mediated models, we can make do with less structure than the O'Raifeartaigh models; higher order supergravity and Kahler potential corrections can stabilize $X$, without additional
fields like $A$.  With
\beq
W = -{1 \over 4} W_\alpha^2 (g^{-2} + c X)
\eeq
we have a Polonyi-type model.  If we simply define $X=0$ as the location of the minimum of the potential, we can expand the Kahler potential about this
point, and impose the conditions of a stable minimum at a the origin with (nearly) vanishing $V$\cite{bosedine}.  Note, in particular, that $X$ is neutral under the $R$
symmetry, so the origin is not a distinguished point.

If we take the gravitino mass to be of order $10$ TeV, we expect stop masses of this order, and can really account for the apparent observed Higgs mass.
But such a choice leaves several questions.
\begin{enumerate}
\item  Raising the scale ameliorates, but does not resolve, the problems of flavor of supergravity models.
This has lead to the suggestion, in \cite{minisplit}, that the scale of supersymmetry breaking should be much
higher, even 1000's of TeV.  Alternatively, one might invoke some model for flavor, e.g. those of \cite{nirseiberg}.
(Other aspects of these question are under study\cite{dds}.
For $10$ TeV squarks, such models are easily compatible with existing data on flavor-changing processes.
\item  $10$ TeV represents a significant tuning.  Even allowing, say, anthropic selection among approximately 
supersymmetric states in a landscape, where might
such a little hierarchy come from?   In this subsection, we will offer a possibility.  Others have been suggested in \cite{yanagidahiggs,dinepack}.
\item  Are there observable consequences of such a picture?  The authors of \cite{minisplit} invoke unification and dark matter to argue
that some gauginos should be relatively light.  In \cite{bosedine}, however, the genericity of light gauginos was questioned.     
\end{enumerate}

Once one has allowed for the possibility that there may be some degree of tuning, the question which immediately follows is:  how much
tuning is reasonable.  A part in $10^3-10^4$?  This would lead to squarks in the $3-10$ TeV range.  A part in $10^6-10^7$?  This would
allow squarks in the $10^3-10^4$ TeV range.  Here we suggest one possible origin for tuning, which points towards the former.

Suppose, for the moment, that we take the $R$ breaking sector to be a pure gauge theory, and we require vanishing of the cosmological constant.  Then we have,
as parameters, the choice of gauge group, the value of the 
gauge coupling at some fixed large scale,
 and a small number of order one terms in the Kahler potential.  Up to order one numbers, the choice of gauge group and the value of the coupling
fix $m_{3/2}$.  We can ask whether we can achieve, among possible groups, $m_{3/2} \approx M_Z$.  To make sense of this question, we need to make
further assumptions.  We will assume that {\it all} of the gauge couplings unify at $M_p$, and employ the standard results for unification within the MSSM.
Then, given a choice of gauge group in the $R$ breaking sector, the scale of that sector, {\it and the value of the gravitino mass}, $m_{3/2}$, are determined.
Confining our attention, for simplicity, to $SU(N)$ theories, 
we have that
\beq
\Lambda = M_{p} e^{- {1 \over b_0}{8 \pi^2 \over g^2(M_p)}}
\eeq
and
\beq
m_{3/2} = {N\Lambda^3 \over M_p^2}
\eeq
For $N$ such that $b_0=3N$ gives a gravitino mass in the TeV range, a change in $N$ by $1$ results in a change in the gravitino mass of order $10^4$.
So, accounting for threshold and other effects, one would expect, typically, to have a graviton mass of order $100$ times $M_Z$ (or $.01~M_Z$).  This might well
account for the sort of tuning needed to account for the Higgs mass, and not much more!  This is, of course, just one possible model; other models might make significantly
different predictions.

\section{Retrofitting Gauge Mediation:  Spontaneous (Continuous) R Symmetry Breaking}
\label{spontaneous}

In broad classes of O'Raifeartaigh models, one finds that the (continuous) $R$ symmetry is unbroken at the minimum of the potential when one performs the requisite Coleman-Weinberg calculation.  In retrofitting such models, and in building gauge-mediated
theories, we need to explicitly break the symmetry, or to insure that there is no such symmetry in the messenger sector.
Instead, in this section, we consider retrofitting in models in which the $R$ symmetry is spontaneously broken.  The simplest such model has superpotential\cite{Shih:2007av}:
\beq
W = X(\phi_1 \phi_{-1} - f) + m_1 \phi_1 \phi_1 + m_2 \phi_{-3} \phi_1
\eeq
We have not explicitly indicated dimensionless couplings.  This model has a metastable minimum at $X\sim m_1,m_2$, provided
\beq
\vert f \vert < \vert m_1 m_2 \vert
\label{shihinequality}
\eeq
When this bound is not satisfied, the model exhibits runaway behavior.
When it is, $F_X=f$ is the order parameter for supersymmetry breaking.

Given these remarks, and the constraint of the cosmological constant, the only possibilities for  retrofitting are
\begin{enumerate}
\item  Comparable $m_1,m_2$:
\beq
f \rightarrow {W_\alpha^2 \over M_p},~{S^3 \over M_p}; ~m_1,~m_2 \rightarrow S
\eeq
with coefficients of order one.
\item
Hierarchy of $m_1,m_2$:
\beq
f \rightarrow {W_\alpha^2 \over M_p},~{S^3 \over M_p}; ~m_1 \sim S,~m_2 \sim {S ^2 \over M_p}
\eeq
or
\beq
f \rightarrow {W_\alpha^2 \over M_p},~{S^3 \over M_p}; ~m_1 \sim {S^2 \over M_p},~m_2 \sim S
\eeq
with suitable order one constants, in each case, so that eqn. \ref{shihinequality} is satisfied.
\end{enumerate}

The latter case, however, is problematic if there are no very small dimensionless numbers.  First, unless $m_1 \gg m_2$,
the $R$ symmetry is unbroken\cite{Shih:2007av}.  Following the analysis of \cite{Shih:2007av}, if this condition is satisfied, the vev of $X$ is:
\beq
\vert \langle X \rangle \vert^2 \approx {m_1^2 \over 9 \lambda^2} \sim \Lambda^2.
\eeq
if the couplings in the superpotential are of order one.  So
the scalar component of $X$ is of order $\Lambda$ (up to dimensionless constants), as in the previous case.

\subsection{Couplings to Messengers}

In the first case, if we couple $X$ to messengers, with
coupling
\beq
X \widetilde{M} M
\eeq
we have the usual sorts of gauge-mediated relations, but with scales that are now, essentially, fixed.
In particular, the scale that sets the masses of squarks, leptons and gauginos is:
\beq
\Lambda_{gm} = {F_X \over X} = {\Lambda^2 \over M_p}
\eeq
(up to dimensionless coupling constants).  Requiring
\beq
\Lambda_{gm} = 10^6 {\rm GeV}
\eeq
(consistent with current experimental constraints, but, needless to say, demanding significant tuning)
gives
\beq
\Lambda = 10^{12} ~{\rm GeV};~~~~m_{3/2} \sim 1 ~{\rm GeV}.
\eeq

The scales here are close to those considered in \cite{fengetal}, who have discussed some of the issues associated with possible detection and
dark matter.  These will be further considered elsewhere, but it should be noted that the lightest of the new supersymmetric particles are in the TeV range,
and these do not carry color, so their discovery will be challenging, if these ideas are correct.

\subsection{The R Axion}

Models of this type, where the approximate $R$ symmetry is spontaneously broken, possess an $R$ axion.  To determine its mass, we
must examine sources of $R$ symmetry breaking.  These will arise from higher dimension terms in the superpotential, and also from coupling the low dimension terms to supergravity.  These latter are always present, so we content ourselves with estimating these.

As in the estimate of Bagger, Poppitz and Randall\cite{bpr}, the $R$ breaking arises from terms such as
$
- 3 \vert W \vert^2$
in the potential.  For the retrofitted versions of Shih's model, writing
\beq
X \approx \langle X \rangle e^{i a/\langle X\rangle}
\eeq
yields a mass of order
\beq
m_a^2 \approx m_{3/2}{f \over X}
\eeq
or about $1$ TeV, in the present case.  This is heavy enough so as not to be astrophysicaly problematic, and, of course, is difficult
to see in accelerator experiments.

\subsection{Discrete Tunings}

In the gravity mediated case, we saw that, with a model for unification of couplings, discrete changes of theory lead to large changes in $m_{3/2}$.  This arose,
in part, because we assumed the simplest possibility for gaugino condensation:  a gauge theory without matter fields.  In the gauge-mediated case, we require
a theory with matter, and, while this may represent an increase in complication, smaller steps in the beta function (one instead of three for the 
pure gauge case) are inherent to
this class of models.  As a result, the difficulties of tuning do not appear to be as pronounced as in the gravity mediated case we described earlier. 
 A ``natural" model of gauge
mediation would have
\beq
\Lambda_{gm} \sim \Lambda_{gm}^{natural} \equiv 4 \times 10^4 {\rm GeV}.
\eeq
If we take the $R$-breaking sector to be an SU(N) gauge theory
with $N_f$ flavors and no particularly small dimensional parameters and makes the same
unification assumptions we made in the gravity-mediated case, it is easy to choose the number of flavors and colors,
so as to obtain $\Lambda_{gm}$ within a factor of three of $\Lambda_{gm}^{natural}$.  So if nature is gauge mediated,
understanding the little hierarchy will require additional elements.  For example, if there is an underlying landscape, and $N$ and $N_f$
are not uniformly distributed, one might easily account for a hierarchy of several orders of magnitude.

\section{Retrofitted Gauge Mediation:  Explicit R Symmetry Breaking}
\label{explicit}

Given the seemingly unlimited ability to introduce scales through retrofitting, one is led to consider models in which
the O'Raifeartaigh sector has an approximate, unbroken continuous $R$ symmetry, while the
would-be $R$ symmetry of the messenger sector is broken by explicit mass terms or couplings in the superpotential.
This would be interesting in itself, but especially because, even with the simplest messenger structure, the spectrum would
be that of general gauge mediation (as opposed to MGM).
But, as we will see in this section, this possibility is remarkably constrained.  It is difficult to construct
realistic models, without very small dimensionless parameters, subject to the following rules:
\begin{enumerate}
\item  $M_p$ sets the overall energy scale of the theory.
\item  The cosmological constant should vanish at the level of the dynamics responsible for supersymmetry
breaking.
\end{enumerate}

A simple model illustrates the main issue.
We consider a retrofitted O'Raifeartaigh model with a field, $X$, neutral under the $R$ symmetry
and with $F$-component
$\Lambda^3/M_p$.  For the coupling to the messengers we take
\beq
\left(y X {S^m \over M_p^m} + \lambda{S^m \over M_p^{m-1}}\right)\widetilde{M}M
\eeq
The problem is that, for any choice of $m$,
\beq
m_{3/2} \approx {\lambda \over y} \Lambda_{gm} 
\eeq
If $\Lambda_{gm} \approx 10^6$ GeV, it is necessary that $\lambda \over y$ be quite small if the gauge-mediated
contributions are to dominate.

The difficulty here arises because $X$ is invariant under the symmetry.  One might try to avoid this by considering
a different type of O'Raifeartaigh model, in which $\vert f \vert \gg \vert m \vert^2$.  For example,
\beq
W = X \left ({S^{2m} \over M_p^{2m-2}} - A^2 \right ) + {S^n \over M_p^{n-1}} AY.
\eeq
If $m<n$, $A$ acquires a vev, and
\beq
F_Y \approx {S^{m+n} \over M_p^{m+n -2}}.
\eeq
Requiring vanishing of the cosmological constant gives
\beq
m+n =3.
\eeq
So there are a limited set of possibilities; indeed, we need
$n=2,m=1$.  But if the fields $S$ transform with $\alpha^{2/3}$ under
discrete R-symmetry, then $Y$ is again neutral, and we encounter
exactly the difficulty of the previous model.

Given these difficulties, one might try to construct a model in which $X$ transforms non-trivially
under the $R$ symmetry.   In a model like
\begin{equation}
  \label{eq:r-sector}
  W= y_f {S^k\over M_p^{k-1}}\widetilde{Q}_fQ_f-{\gamma\over p}{S^p\over M_p^{p-3}}
\end{equation}
$S$ transforms as $\alpha^{2/p}$.  But now if we are to replicate our ``cosmological constant
coincidence", we require that $X$ couple to $S^p \over M_p^{p-4}$.  But then $X$ is neutral
again.


There are other strategies one might try, but it seems difficult, in
general, to break the $R$ symmetry subject to our rules.  Needless to
say, relaxing these would open up additional possibilities.

\section{Explicit R Breaking By Supergravity}
\label{supergravity}

Finally, one might wonder whether simply coupling one of these systems to supergravity might provide an adequate
breaking of the continuous $R$ symmetry\footnote{Supergravity corrections of this type in gauge mediation have been
considered by Kitano\cite{kitano}.}.

In the simplest OR model, coupled to messengers:
\beq
W = X f + \lambda X A^2 + mAY + c ~f~ M_p | \gamma X M \bar M
.\eeq
(with $c$ an ${\cal O}(1)$ constant),
the tadpole (linear term in the potential) for $X$ is of order
\beq\Gamma \approx {f^2 \over M_p}.~~~~~
m_X^2 = {\lambda^4~f^2 \over 16 \pi^2 m^2}.\eeq
So, if $f \sim {\Lambda^3 \over M_p}$ and $m \sim \Lambda$,
\beq X \approx {\Gamma \over m_X^2} \sim {\Lambda^2 \over M_p} \left ( {\lambda^4 \over 16 \pi^2} \right )^{-1} 
\eeq

The simplest coupling to messengers again has the MGM form:
\beq
\gamma X M \bar M
\eeq
There are now two conditions on $\gamma$ and $\lambda$.  First, we require that the messenger masses not be tachyonic:
\beq
\vert \gamma X \vert > \vert F_X \vert
\eeq
and second that the corrections to the $X$ potential due to the messengers be small compared to those from the $X$ interactions with
the massive field $A$:
\beq\label{eq:weakmessenger}
{\gamma^2 \over X^2} \ll {\lambda^4 \over m^2}.
\eeq
These conditions require that both $\lambda$ and $\gamma$ be small, but they do not have to be {\it extremely} small.
For example, they are satisfied with
\beq
\lambda = 0.08;  ~~~\gamma = 0.01; ~~ \gamma X \approx 10^{12}~\textrm{GeV}.
\eeq
A slightly smaller $\lambda$ yields $X$ at the maximum scale for gauge mediation, while allowing a larger $\gamma$:
\beq
\lambda = .05;~~~~\gamma = 0.10; ~~~\gamma X \approx 10^{15}~\textrm{GeV}.
\eeq
On the other hand, once $\lambda$ is larger than about $0.18$, $\gamma$ becomes non-perturbatively large.

So overall, one can achieve a realistic model in this manner, with $\lambda$ and $\gamma$ which are small but not
extremely so.  The gauge mediated scale can range over the full range normally considered for gauge mediated
models; the simplest models have the spectrum of MGM.

\section{Conclusions}
\label{conclusions}

It seems likely that our cherished ideas about naturalness and supersymmetry are not correct.  Supersymmetry, if present at low energies,
appears somewhat tuned and may be hard, or impossible, to find.  The apparent value of the Higgs mass suggests that
the supersymmetry breaking scale might be in the $10-100$ TeV range.  

In this paper, we have reexamined the question of dynamical supersymmetry breaking in the framework of retrofitted models.  These models
appear to have a rather generic character, and allow one to address easily questions ranging from the $\mu$ term to the cosmological
constant.  With plausible assumptions, they are highly constrained.  We have considered gravity mediated models (extending slightly the
work of \cite{bosedine}) and gauge mediated models.  In both cases, the requirement of small cosmological constant strongly constrains the underlying theory.  In the supergravity case, the question of fine tuning, i.e. of how close $m_{3/2}$ lies to $M_Z$, is a question
of discrete choices.  With plausible assumptions about the microscopic theory, the apparent degree of tuning is typically a part in thousands or
tens of thousands, perhaps explaining the tuning we see.  It is still necessary, in this case, that there be some suppression of low energy
flavor violation.  Models along the lines of \cite{nirseiberg} which achieve this will be considered elsewhere.

Our principle focus, however, was on gauge mediated models.  We constrained our constructions, again, by requiring the possibility of small
cosmological constant in the effective theory,
and a fixed supersymmetry
breaking scale (corresponding to stops at $10$ TeV, or $\Lambda_{gm} = 10^6$ GeV).   
We explored the question of whether one might break the approximate, continuous $R$ symmetry explicitly, taking advantage of the
freedom apparently implied by the retrofitted constructions.  While we cannot claim that our survey of possible constructions are complete,
in broad classes of theories:
\begin{enumerate}
\item  If the $R$ symmetry is spontaneously broken, and absent very small dimensionless couplings, the underlying scale of supersymmetry
breaking is fixed, with a gravitino mass of order $1$ GeV.
\item  If the $R$ symmetry is explicitly broken through retrofitted couplings in the superpotential, a very small dimensionless number, of order $10^{-6}$, is required in order that the
gauge-mediated contributions dominate.
\item  If the $R$ symmetry is explicitly broken by supergravity effects, two small, but not exceptionally small couplings, are required.  The
has scale of the messengers ranges, in simple cases, from $10^{7}$ to $10^{15}$ GeV.
\end{enumerate}

We draw from these observations the conclusions:
\begin{enumerate}
\item  If supersymmetry breaking is gravity mediated, the relatively high scale may result from the limited effectiveness of required discrete
tuning.  Flavor symmetries, associated with quark and lepton masses, readily can provide adequate alignment of soft breakings to
suppress low energy flavor changing processes\cite{ads}.
\item  If supersymmetry breaking is gauge mediated, the approximate $R$ symmetry may be spontaneously broken, in which
case the underlying scale of supersymmetry breaking corresponds to a gravitino mass of order $1$ GeV, and the mass of the
corresponding $R$ axion is similar.  Simple models of Minimal Gauge Mediation can be realized in this framework.
\item  The breaking may be explicit.  In the most compelling models, the breaking of the $R$ symmetry arises from
supergravity effects.  The messenger scale may be small or large, and again MGM can be realized.
\end{enumerate}

There remains the most important question:  is there anything one might hope to see\cite{fengetal}.  In a subsequent publication, we will focus on this issue, considering
questions such as dark matter and its implications for possible light states, electric dipole moments, and rare processes.

\noindent
{\bf Acknowledgements:}
This work supported in part by the U.S.
Department of Energy.

\bibliographystyle{unsrt}
\bibliography{dinerefs}{}

\end{document}